\documentclass[conference]{IEEEtran}

\IEEEoverridecommandlockouts

\usepackage{cite}
\usepackage{amsmath,amssymb,amsfonts}
\usepackage{algorithmic}
\usepackage{graphicx}
\usepackage{textcomp}
\usepackage{xcolor}

\usepackage{balance}
\usepackage[hidelinks]{hyperref}
\usepackage{cleveref}

\begin{document}

\title{Real-World Chaos-Based Cryptography Using Synchronised Chua Chaotic Circuits\\
\thanks{Funded by DFG, under Projects 440182124 and 439892735 of SPP 2253, and by BMBF under Joint Project 16KISK034 for the 6G-RIC.
}
}

\author{\IEEEauthorblockN{Emiliia Nazarenko\IEEEauthorrefmark{1}, Nikolaos Athanasios Anagnostopoulos\IEEEauthorrefmark{1}\IEEEauthorrefmark{2}, Stavros G. Stavrinides\IEEEauthorrefmark{3},\\Nico Mexis\IEEEauthorrefmark{1}, Florian Frank\IEEEauthorrefmark{1}, Tolga Arul\IEEEauthorrefmark{1}\IEEEauthorrefmark{2}, Stefan Katzenbeisser\IEEEauthorrefmark{1}}

\IEEEauthorblockA{\IEEEauthorrefmark{1}Faculty of Computer Science and Mathematics, University of Passau, Innstraße 43, 94032 Passau, Germany\\
Emails: \{nazare02, anagno02, mexis01, frank55, arul01, katzen07\}@ads.uni-passau.de\\}
\IEEEauthorblockA{\IEEEauthorrefmark{2}Computer Science Department, Technical University of Darmstadt, Hochschulstraße 10, 64289 Darmstadt, Germany\\
Emails: \{anagnostopoulos, arul\}@seceng.informatik.tu-darmstadt.de}
\IEEEauthorblockA{\IEEEauthorrefmark{3}School of Science and Technology, International Hellenic University, Thermi Campus, 57001 Thessaloniki, Greece\\
Email: s.stavrinides@ihu.edu.gr}
}

\maketitle

\begin{abstract}
This work presents the hardware demonstrator of a secure encryption system based on synchronised Chua chaotic circuits. In particular, the presented encryption system comprises two Chua circuits that are synchronised using a dedicated bidirectional synchronisation line. One of them forms part of the transmitter, while the other of the receiver. Both circuits are tuned to operate in a chaotic mode. The output (chaotic) signal of the first circuit (transmitter) is digitised and then combined with the message to be encrypted, through an XOR gate. The second Chua circuit (receiver) is used for the decryption; the output chaotic signal of this circuit is similarly digitised and combined with the encrypted message to retrieve the original message. Our hardware demonstrator proves that this method can be used in order to provide extremely lightweight real-world, chaos-based cryptographic solutions.
\end{abstract}


\section*{Introduction}

The rapid increase in the number of electronic devices in everyday use leads to an unlimited growth of vulnerable communication that must be protected from possible attacks. Having in focus the growing Internet of Things ecosystem, as well as edge computing, secure data transmission appears as a very important part. One of the strongest tools for mitigating attacks on electronic communications, is cryptography.

In our work, we focus on securing the transmitted information through symmetric stream cryptography that is based on the chaotic signal produced by two synchronised Chua chaotic circuits~\cite{Nikolaus}, one of which forms part of the transmitter and the other of the receiver. The produced chaotic signal acts as a random number stream that is shared by the synchronised chaotic circuits. Thus, the message can be encrypted at the transmitter by being XORed with a digitised form of the aforementioned chaotic signal, and decrypted at the receiver by XORing the encrypted message stream with the digitised form of the same chaotic signal.

\section*{Description of the Hardware Demonstrator }

\begin{figure}[t]
\centerline{\includegraphics[width=0.8\linewidth]{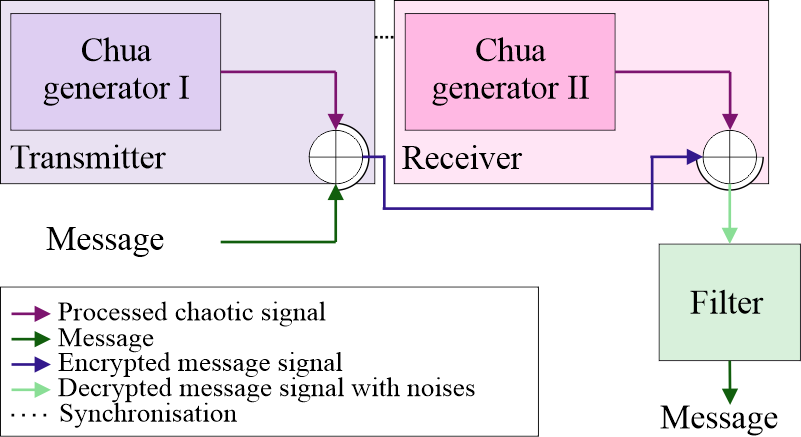}}
\vspace{-5pt}
\caption{The overall architecture of the encryption-decryption system.}
\label{img:Architecture:Schema}
\vspace{-15pt}
\end{figure}

\begin{figure}[!b]
\vspace{-20pt}
\centerline{\includegraphics[width=0.70625\linewidth]{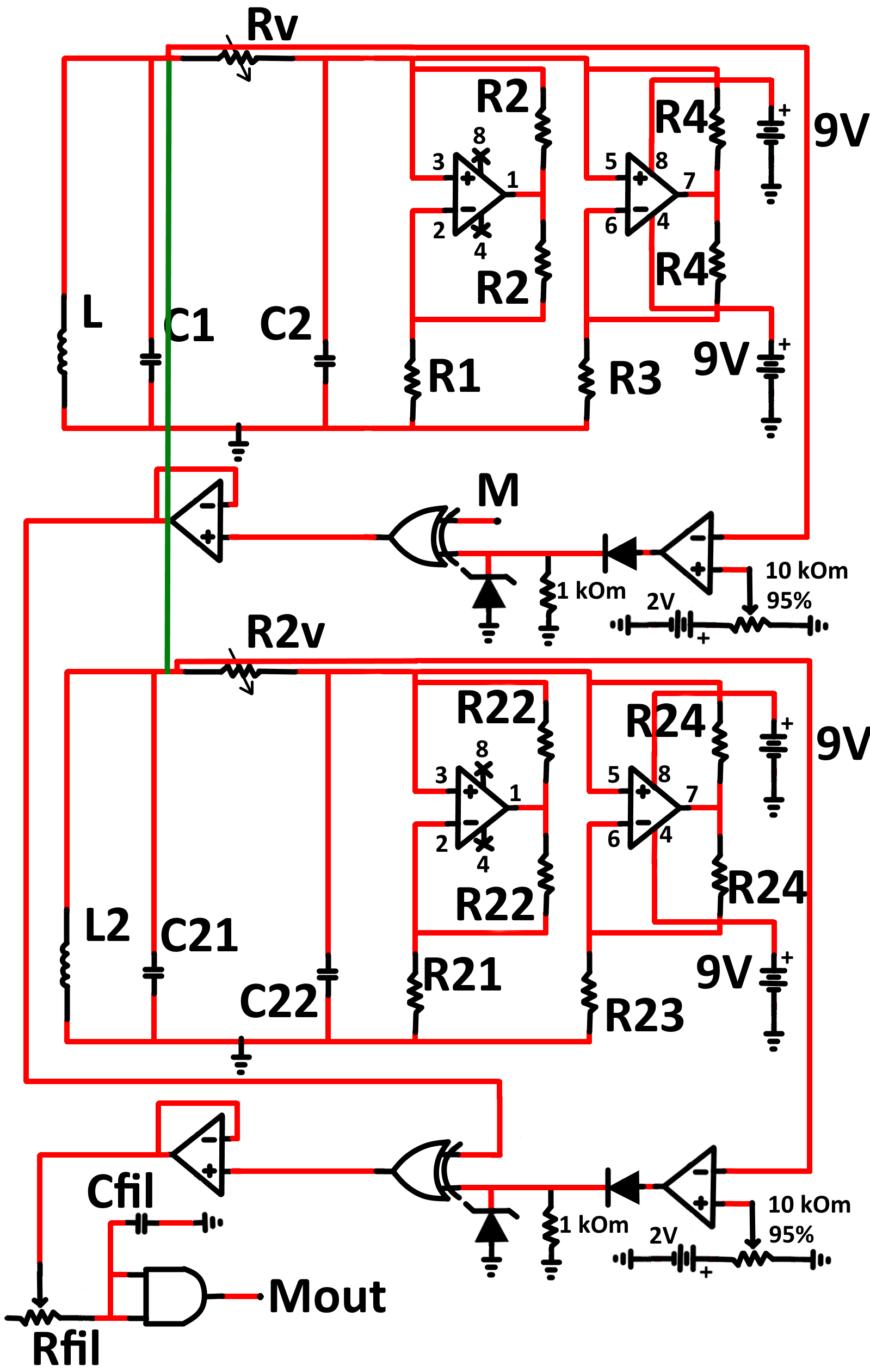}}
\vspace{-5pt}
\caption{Schematic diagram of the overall implemented system.}
\label{img:AllSystem}
\end{figure}

A simplified architecture of the proposed synchronized chaotic encryption-decryption system is shown in \Cref{img:Architecture:Schema}. A proof-of-concept circuit of this system has been designed and appears in \Cref{img:AllSystem}, which demonstrates the full circuitry. The relevant characteristics and values of the components used for implementing the circuits are listed in \Cref{Arch:Table:ChuaComponetns}. 

The Chua circuits that we are using, are based on the ones examined by Kennedy~\cite{Kennedy+1992+66+80}. In our approach, two channels are utilized, one for achieving synchronisation and another for transmitting information. The signal produced by the chaotic circuits is digitised and processed according to the method described in~\cite{stav2, svt3}.
After being processed, this signal is used to encrypt a message, by performing an XOR operation. In order to perform a successful decoding, the transmitter and receiver are synchronised using a dedicated bidirectional synchronisation line. 

\begin{figure}[b]
\vspace{-12.5pt}
\centerline{\includegraphics[width=\linewidth]{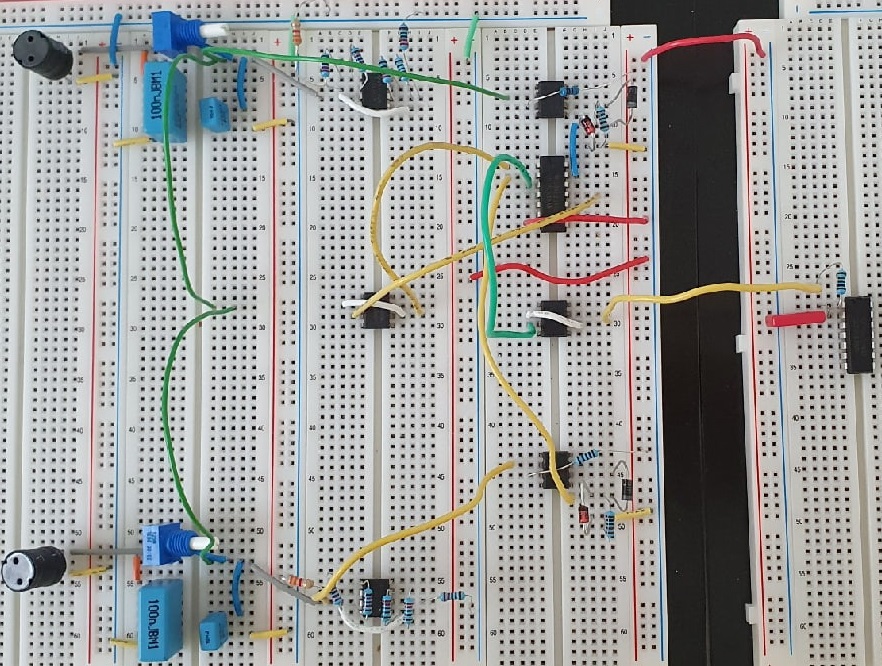}}
\vspace{-5pt}
\caption{The implementation of the proposed system.}
\label{Photo}
\end{figure}

In our demo, the message is generated by a wave generator with the following parameters: frequency $6$\,kHz, amplitude $2.5$\,Vpp, offset $+1.25$\,V, phase $0.0^\circ$, and duty cycle $50\%$. The encrypted message (coming from the XOR) is transmitted to the receiver, which in its turn decodes it, using the XOR operation and its own synchronized chaotic signal, which is processed in a similar way as the chaotic signal of the transmitter. The decrypted message is then cleared by an RC low-pass filter. A real-world implementation of the system is shown in \Cref{Photo}.

\begin{table}[t]
\centering
\caption{Components of the Chua Circuit and Their Characteristics}
\label{Arch:Table:ChuaComponetns}
\vspace{-5pt}
\renewcommand*{\arraystretch}{1.2}
\begin{tabular}{| c | c |} 
\hline
\textbf{Component} & \textbf{Characteristics}\\ 
\hline
Inductors $L, L2$ & \multicolumn{1}{|c |}{18\,mH, 10\% tolerance}\\ 
\hline
Capacitors $C1, C21$ &  \multicolumn{1}{|c |}{100\,nF, 5\% tolerance} \\ 
\hline
Capacitors $C2, C22$ &   \multicolumn{1}{|c |}{10\,nF, 5\% tolerance} \\ 
\hline
Variable resistors $Rv, R2v$ &   \multicolumn{1}{|c |}{1555\,$\Omega$} \\
\hline
Resistors $R1 ,R21$ &   \multicolumn{1}{|c |}{3.3\,k$\Omega$, 5\% tolerance} \\ 
\hline
Resistors $R2, R22$ &   \multicolumn{1}{|c |}{22\,k$\Omega$, 5\% tolerance} \\ 
\hline
Resistors $R3, R23$ &   \multicolumn{1}{|c |}{2.2\,k$\Omega$, 5\% tolerance} \\ 
\hline
Resistors $R4, R24$ &    \multicolumn{1}{|c |}{220\,$\Omega$, 5\% tolerance} \\
\hline
OpAmps  &    \multicolumn{1}{|c |}{TL082ACP, TL081DIP} \\ 
\hline
Batteries &   \multicolumn{1}{|c |}{2\,V, 9\,V} \\ 
\hline
Capacitor $Cfil$ &   \multicolumn{1}{|c |}{7\,nF, 5\% tolerance} \\ 
\hline
Resistor $Rfil$ &   \multicolumn{1}{|c |}{1\,k$\Omega$, 5\% tolerance} \\ 
\hline
\end{tabular}
\vspace{-15pt}
\end{table}

\begin{figure}[htb!]
\vspace{-5.5pt}
\centerline{\includegraphics[width=\linewidth]{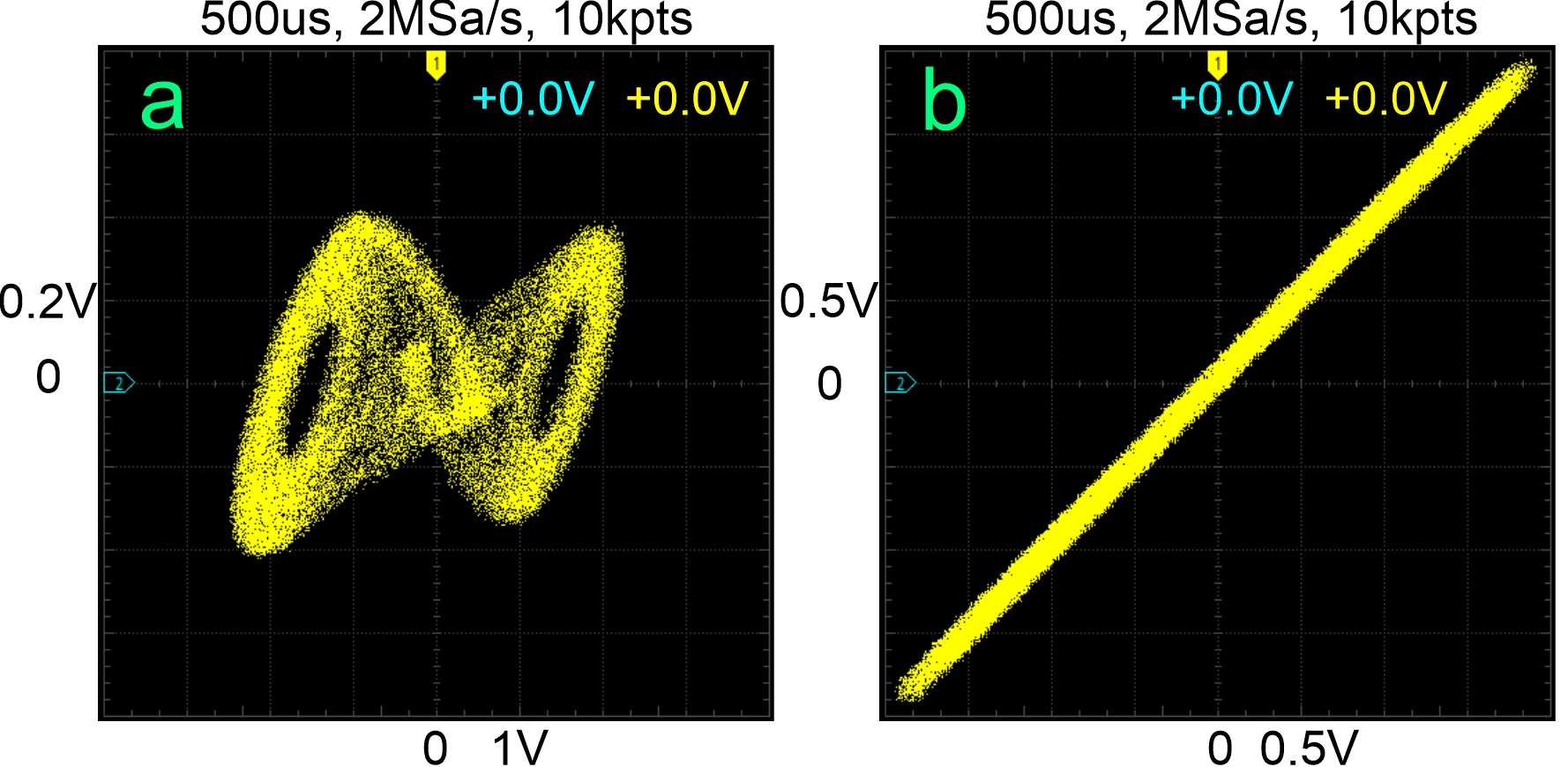}}
\vspace{-7.5pt}
\caption{(a) Phase portrait of the transmitter Chua circuit, coming from the voltages on the capacitors C1 and C2. (b) Synchronisation phase portrait between the chaotic signals of the transmitter and the receiver.}
\label{phase}
\vspace{-15pt}
\end{figure}

\begin{figure}[htb!]
\centerline{\includegraphics[width=\linewidth]{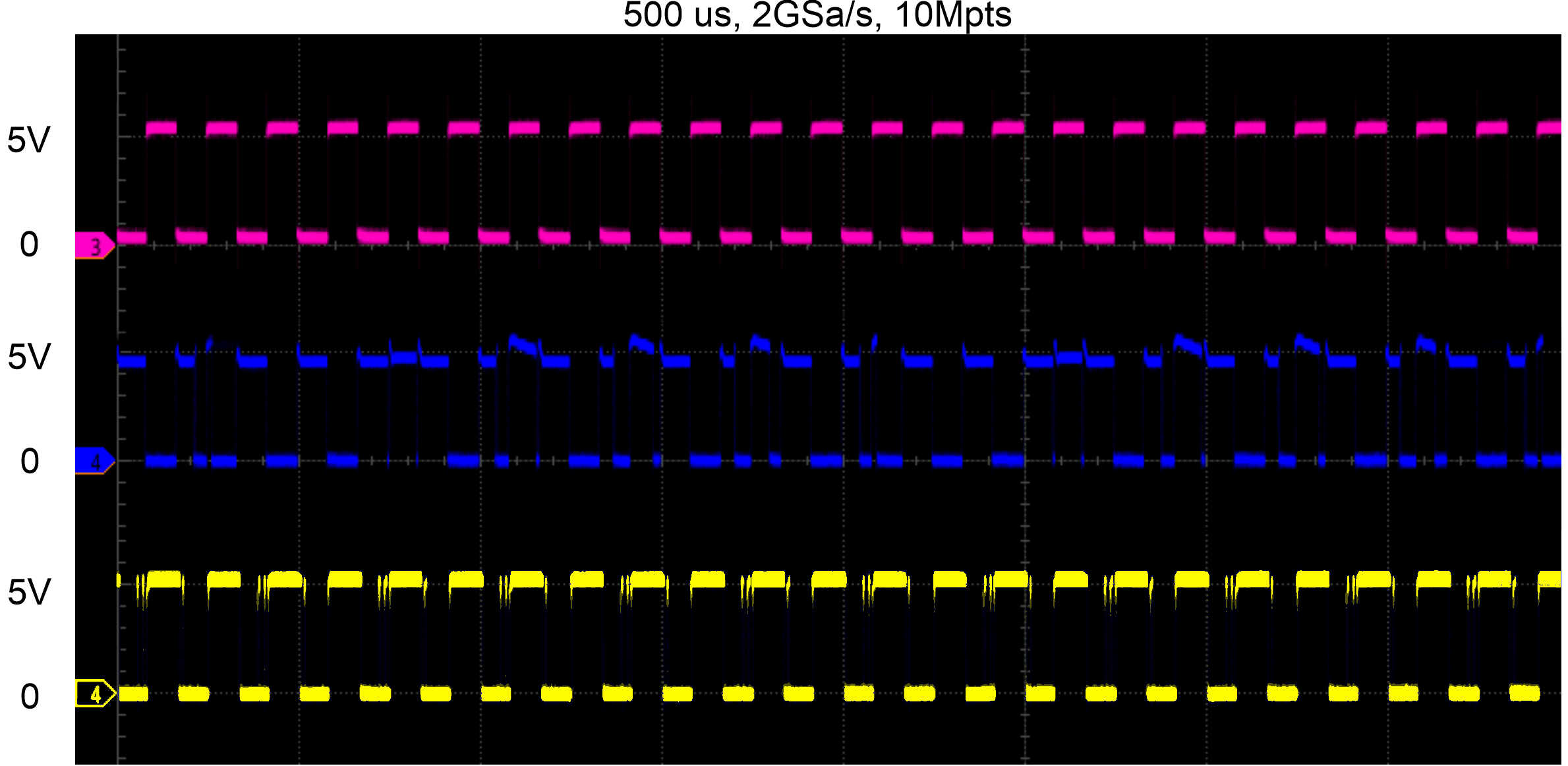}}
\vspace{-7.5pt}
\caption{Initial message (magenta), encrypted message (blue) and decrypted message (yellow). The coincidence between the initial and the decrypted message is evident.}
\label{init}
\vspace{-15pt}
\end{figure}

In \Cref{phase}a, the attractor produced by the transmitter's circuit is illustrated. The chaotic mode of operation of the Chua circuit is evident. The transmitter and the receiver are reliably chaotic-synchronised, as demonstrated by the synchronization phase portrait appearing in \Cref{phase}b, based on the voltages on capacitors C1 and C21. The synchronization quality (perfect in this case) is responsible for the system's ability to provide  efficient decryption. Finally, \Cref{init} presents the initial message (magenta) encrypted in the transmitter, the encrypted message (blue) that is transmitted, and the decrypted message (yellow) at the receiver's output. It is rather evident that the initial message and the decrypted one match.

The circuitry shown in \Cref{Photo}, together with the relevant power supply (batteries), oscilloscope, and wave generator, form the main part of our hardware demonstrator setup.

\section*{Conclusion}

Concluding, one may support that our hardware demonstrator proves that the proposed encryption-decryption system, which is based on two synchronized Chua chaotic circuits, can provide an efficient, lightweight, and practical solution for real-world cryptographic applications.

\bibliographystyle{./IEEEtran}
\bibliography{./IEEEabrv,./IEEEexample}

\end{document}